\documentstyle[12pt]{article}

\large
\newcommand{\be}{\begin{eqnarray}}
\newcommand{\ee}{\end{eqnarray}}
\newcommand\del{\partial}

\begin{document}
\setlength{\baselineskip}{21pt}
\pagestyle{empty}
\vfill
\eject
\begin{flushright}
SUNY-NTG-95/3
\end{flushright}

\vskip 2.0cm
\centerline{\Large Universal fluctuations in
spectra of the lattice Dirac operator}
\vskip 2.0 cm
\centerline{\bf M.A. Halasz and J.J.M. Verbaarschot}
\vskip .2cm
\centerline{Department of Physics}
\centerline{SUNY, Stony Brook, New York 11794}
\vskip 2cm

\centerline{\bf Abstract}
Recently, Kalkreuter obtained the complete Dirac spectrum for an
$SU(2)$ lattice gauge theory.
We performed a statistical analysis
of his data and found that the eigenvalue correlations can be described
by the Gaussian Symplectic Ensemble. Long range fluctuations are strongly
suppressed: the variance
of a sequence of levels containing $n$ eigenvalues on average
is given by $\frac 1{2\pi^2}(\log n + {\rm const.})$.
Our findings are
in agreement with the anti-unitary symmetry of the lattice
Dirac operator for $N_c=2$ with staggered fermions. For $N_c = 3$  we
predict that the eigenvalue correlations are given by the Gaussian Unitary
Ensemble.

\vfill
\noindent
\begin{flushleft}
January 1995
\end{flushleft}
\eject
\pagestyle{plain}

The QCD Dirac operator is of fundamental importance for the calculation
of the physical properties of QCD. Knowledge of its
eigenvalues and eigenfunctions determines the propagator, a necessary
ingredient for the calculation of hadronic correlation functions.
In this letter we will
focus on the eigenvalues of the Dirac operator and isolate some universal
properties that can be understood from its symmetries only.
One such example is already
known: the eigenvalues near zero satisfy sum-rules \cite{LS} with a
generating function given by random matrix theories with the global
symmetries of QCD \cite{V}. This  led us to the conjecture that the
fluctuations of the eigenvalues
no more than a few level spacings away from zero,
over the ensemble of gauge field configurations are given by universal
functions that can be obtained from a much simpler
random matrix theory as well. This
raises the question whether the eigenvalues in the bulk of the spectrum
show such universal characteristics as well.

Recently, in a ground breaking work, Kalkreuter \cite{Kalkreuter} succeeded to
compute the complete spectrum of the Dirac operator on
a reasonable large lattice ($12^4$). His results were in complete
agreement with an analytical sum rule adding greatly to our
confidence in the accuracy of his results.
Long level sequences have been analyzed before in atomic and nuclear
physics and for systems with only a few degrees of freedom \cite{Koch,Bohigas}.
Generically, it was found that if the system is classically
chaotic, the correlations between eigenvalues with the same exact
quantum numbers are given by the gaussian random matrix ensembles.

In this letter we will perform a statistical analysis of the lattice
spectra using such methods \cite{Bohigas}.
We start from the assumption that the average eigenvalue
density, $\bar \rho(\lambda)$, can be separated from the fluctuations of
the eigenvalues about their average position.
This allows us to unfold the spectrum. This
is a procedure in which the eigenvalues are rescaled according to
the average local level density. Formally, the unfolded spectrum
$\{ { \lambda'}_n\}$, with average eigenvalue density $\bar\rho'(\lambda)=1$,
is given by
\be
\int_{-\infty}^{\lambda_n} \bar \rho(\lambda) d\lambda = \lambda'_n.
\ee

{}From the analysis of the spectra of the Hamiltonians
of classically chaotic systems, we have
learned that depending on the time reversal symmetry of the system
the level correlations fall into three different universality
classes. We want to stress that an $anti$-$unitary$ symmetry
determines whether the matrix elements
are complex, real or quaternion real \cite{DYSON-three}. The corresponding
invariant random matrix ensembles
are called the Gaussian Unitary Ensemble (GUE),
the Gaussian Orthogonal Ensemble (GOE) and the Gaussian Symplectic Ensemble
(GSE), respectively.
According to general universality arguments \cite{BZ} the
correlations between eigenvalues in the $bulk$ of the spectrum, as
opposed to those near the edge of the spectrum, are not sensitive to
many other details of the random matrix model. For example,
the  eigenvalue fluctuations in the bulk of the spectrum of random
matrix models \cite{V}, with the chiral symmetry of the Dirac operator
built in,
will be given by one of the invariant random matrix models (GUE, GOE or GSE).

Let us analyze the anti-unitary symmetries of the Euclidean Dirac operator,
${\cal D} \equiv i\gamma_\mu\del_\mu + \gamma_\mu A_\mu$,
for fundamental fermions in an arbitrary $SU(N_c)$ background gauge field
$A_\mu$.
For three or more colors there are no anti-unitary symmetries and the
matrix elements of the Dirac operator ${\cal D}$ are complex.
For two colors the
situation is different. In the continuum theory we have \cite{LS}
\be
[{\cal D}^{\rm cont.}, C\tau_2 K] = 0,
 \ee
where $C$ is the charge conjugation matrix ($C= \gamma_2\gamma_4$), $\tau_2$
is one of the Pauli spin matrices, and $K$ is the charge  conjugation
operator. Naive lattice fermions also obey this symmetry, but for Wilson
fermions it is violated by the $r-$term.
For staggered lattice fermions the
only remnant of the $\gamma-$matrices is a phase factor $\pm 1$ and instead
of (2) we have \cite{Wiese}
\be
[{\cal D}^{\rm stag.}, \tau_2 K]=0.
\ee
The important difference is that
\be
(C\tau_2 K )^2 = 1,\quad {\rm but} \quad(\tau_2 K)^2 = -1.
\ee
{}From a similar analysis of the time reversal operator in  quantum mechanics
(see \cite{Porter})
we conclude that in the continuum theory the matrix elements of the
Dirac operator can be chosen $real$ whereas for staggered fermions
they can be organized into $real$ $quaternions$. Therefore, we expect
that the level correlations of the
eigenvalues of ${\cal D}^{\rm cont.}$  are described by the GOE,
whereas for ${\cal D}^{\rm stag.}$ they are given by the GSE.
A necessary condition in both cases is that the gauge potential
is 'sufficiently random'.
For Wilson fermions
the matrix elements are complex even for two colors, but at present
it is not clear whether breaking of
the anti-unitary symmetry by an irrelevant operator leads to different
level statistics.

In his work, Kalkreuter \cite{Kalkreuter} gives results for only a few
$SU(2)$ gauge field configurations, namely for $\beta = 4/g^2= 1.8$ and
$\beta=2.8$, both with periodic and anti-periodic boundary conditions in the
Euclidean time direction (in the spacial directions he uses periodic boundary
conditions). His configurations
have been obtained for dynamical staggered fermions with a mass of 0.2.
For $\beta =1.8$ the theory is certainly in the strong coupling
phase, but it is generally believed that asymptotic scaling has been
reached for $\beta =2.8$.

With the eigenvalues of only one configuration
we are unable to perform an ensemble
average of their correlations. Instead we will
perform a spectral average, which, at least in random matrix theory
can be shown to be equal to the ensemble average \cite{Pandey}.
For the present data we have
verified numerically that the eigenvalue correlations
do not change over the range  of the spectrum.

The integrated level density (see Fig. 1)
\be
N(\lambda) = \int_0^\lambda \rho(\lambda') d\lambda',
\ee
of the eigenvalues of ${\cal D}^{\rm stag.}$  follows immediately
from the 10368 eigenvalues calculated by Kalkreuter \cite{Kalkreuter}.
Except for a
possible nonanalytical behaviour at the ends of the spectrum, it is
extremely smooth and almost linear. To unfold the spectrum we cut it
in pieces of 500 eigenvalues and fit them by a second order polynomial.
We have checked that this results in an average eigenvalue density equal to
one,
and that our results are insensitive to the details of the cuts.

To define our statistics we introduce the quantity
\be
N(x, x+n) = \int_x^{x+n} \rho'(\lambda) d\lambda,
\ee
where $\rho'(\lambda)$ is the unfolded spectral density. So, $N(x, x+n)$
is the number of eigenvalues in a sequence of length $n$ starting at $x$.
The $number$ $variance$ is defined by
\be
\Sigma_2(n)= \frac 1p\sum_{i=1}^p (N(x_i,x_i+n)-n)^2,
\ee
where the points $x_i$ are regularly spaced such that the sequences
$[x_i,x_i+n]$ cover the complete spectrum.
It can be related to the two-point level correlation function which
is known analytically in random matrix theory (see \cite{Mehta}).
For a random sequence of levels (Poisson spectrum)
it can be  shown that $\Sigma_2(n) = n$ (see \cite{Bohigas}).

Our results for $\Sigma_2(n)$ are shown in the upper figure  of Fig. 2
($\beta =1.8$) and Fig.~3 ($\beta = 2.8$).
(To eliminate inaccuracies in the unfolding procedure we excluded 500
eigenvalues at the ends of the spectrum.)
Long range fluctuations are
almost completely absent. Instead of a variance of 100 for an average
sequence of length 100 we find a variance of only 0.4, showing the presence
of very strong correlations between the eigenvalues.
{}From what we have  said before, we expect that they
can be described by the GSE. Indeed,
the theoretical result (full curve), for $r\ge 1$ given by
\be
\Sigma_2(r) = \frac 1{2\pi^2}(\log(4\pi r) +\gamma + 1 +\frac{\pi^2}8)
+O(\frac 1{\pi^2 r}),
\ee
shows a perfect agreement with the lattice data.
For comparison, we have also given the result for the GUE (dashed curve).
The result for the GOE is for the most
part outside the range of the figure.

A much smoother statistic is the $\Delta_3$-statistic originally introduced
by Dyson and Mehta \cite{DM}.
It is related  to the number variance by \cite{Brody}
\be
\Delta_3(r) = \frac 2{r^4} \int_0^r(r^3 - 2r^2 s +s^3)\Sigma_2(s) ds,
\ee
and does not receive contributions from the quadratic term in $\Sigma_2(s)$
so that small unfolding errors are eliminated.
Asymptotically, for $r\ge 1$, one finds for the GSE
\be
\Delta_3(r) = \frac 12 \Sigma_2(r)- \frac 9{16\pi^2}.
\ee
Results for $\Delta_3$ are shown in the middle figure of Fig. 2 and Fig. 3.
Numerical results obtained from the data of Kalkreuter \cite{Kalkreuter} are
given by full dots whereas the full curve refers to the GSE result
and the dotted curve represents the $\Delta_3$ statistic of the GUE.
Also in this case the curve for the GOE lies well above the GUE.

Finally, in the lower figure of Fig. 2 and Fig. 3,
we show the distribution $P(S)$ of the nearest neighbor
spacings $S =|{\lambda'}_{i+1}-{\lambda'}_i|$ of the unfolded spectrum.
The histograms obtained from the lattice
data are represented by points. The full lines show the GSE result
for the nearest neighbor
spacing distribution and the dashed curve give the result for the GUE.
The exact expression is very well
approximated by the Wigner surmise which for the GSE is given by
\be
P(S) = \frac{2^{18}}{3^6 \pi^3} S^4 \exp\left(-\frac{64}{9\pi} S^2\right ).
\ee
For comparison we also show the result for the GUE (dashed curve) which is
quadratic in $S$ for small $S$. Concerning the spacing distribution of the GOE
we only remark that it starts out linearly and deviates even more from the GSE.
The nearest neighbor spacing distribution
contains information on short range correlations in the spectrum. We hope
that we have convinced the reader that they are also given by the GSE.

We have also analyzed the spectra that Kalkreuter obtained for periodic
boundary conditions in all directions and found identical results for
the eigenvalue correlations.

In conclusion, we have found that the fluctuations of the eigenvalues of
the staggered Euclidean Dirac operator for $SU(2)$-color
can be described
by the Gaussian Symplectic Ensemble both for $\beta=1.8$ and $\beta=2.8$.
For the continuum theory the anti-unitary
symmetry is different and we expect
fluctuations according to the Gaussian Orthogonal Ensemble. In \cite{V,SV}
we have shown that anti-unitary symmetries are essential in determining the
Goldstone sector of the theory. Because the massless sector
for staggered fermions differs from the continuum theory, it is not
surprising to find level correlations that belong to a different
universality class. It would be very interesting to analyze the fate of the
anti-unitary symmetries and eigenvalue correlations
in the continuum limit of the lattice gauge theory.
Lattice simulations with larger values of $\beta$ and  bigger lattices
are required to investigate this point.

For three or more colors the anti-unitary symmetries are broken both in the
continuum theory and for staggered lattice fermions.
We predict that in this case the eigenvalue correlations
are given by the GUE. It would be worthwhile to obtain
the complete Dirac spectrum also for this case.

A final point of interest we want to mention is the fate of
level correlations during the chiral phase transition. From solid state
physics \cite{alt}
we know a delocalization transition is associated with a transition
in the level statistics which raises the hope that such phenomena can be seen
in QCD as well.

\vglue 0.6cm
{\bf \noindent  Acknowledgements \hfil}
\vglue 0.4cm
 The reported work was partially supported by the US DOE grant
DE-FG-88ER40388. We are grateful to Thomas Kalkreuter who made
his eigenvalue spectra available and enabled us to carry out this
investigation.

\vfill
\eject
\newpage
\setlength{\baselineskip}{17pt}

\vfill
\eject
\newpage
\noindent{\bf Figure Captions}
\vskip 0.5 cm
\noindent
Fig. 1. The integrated eigenvalue density $N(\lambda)$ of the Euclidean
lattice Dirac operator for $SU(2)$-color with staggered fermions. Each
curve is the result for one equilibrated gauge field configuration with
$\beta$ as indicated in the figure.

\vskip 0.5cm

\noindent
Fig. 2. Results for the level statistics $\Sigma_2(n)$ (upper),
$\Delta_3(n)$ (middle) and
the nearest neighbor spacing distribution $P(S)$ (lower)
of the eigenvalues shown in Fig. 1 for $\beta =1.8$. The
random matrix results for the GSE and the GOE are represented by the
full and the dashed line, respectively.

\vskip 0.5cm

\noindent
Fig. 3. Results for the level statistics $\Sigma_2(n)$ (upper),
$\Delta_3(n)$ (middle) and
the nearest neighbor spacing distribution $P(S)$ (lower)
for $\beta =2.8$. For further explanation see the caption of Fig. 2.

\end{document}